\documentclass[a4paper,fleqn]{cas-dc}

\usepackage[numbers]{natbib}
\usepackage{subfig}

\begin{document}
\let\WriteBookmarks\relax
\def\floatpagepagefraction{1}
\def\textpagefraction{.001}
\shorttitle{Explainable Depression Detection (MDHAN)}
\shortauthors{xxxxxxxx et~al.}

\title [mode = title]{Explainable Depression Detection with Multi-Modalities Using a Hybrid Deep Learning Model on Social Media}                      



\author[1]{Hamad Zogan}
\address[1]{University of Technology Sydney (UTS), Australia}

\author[2]{Imran Razzak}
\address[2]{Deakin University, Australia}

\author[1]{Xianzhi Wang}
\author[3]{Shoaib Jameel}
\address[3]{University of Essex, United Kingdom}


\author[1]{Guandong Xu}



\begin{abstract}
Model interpretability has become important to engenders appropriate user trust by providing the insight into the model prediction. However, most of the existing machine learning methods provide no interpretability for depression prediction, hence their predictions are obscure to human. In this work, we propose interpretive Multi-Modal Depression Detection with Hierarchical Attention Network \textbf{MDHAN}, for detection depressed users on social media and explain the model prediction. We have considered  user posts along with Twitter-based multi-modal features, specifically, we encode user posts using two levels of attention mechanisms applied at the tweet-level and word-level, calculate each tweet and words' importance, and capture semantic sequence features from the user timelines (posts). Our experiments show that \textbf{MDHAN} outperforms several popular and robust baseline methods, demonstrating the effectiveness of combining deep learning with multi-modal features. We also show that our model helps improve predictive performance when detecting depression in users who are posting messages publicly on social media. \textbf{MDHAN} achieves excellent performance and ensures adequate evidence to explain the prediction.

\end{abstract}




\maketitle

\section{Introduction}
\label{S:1}
Mental illness is a serious issue faced by a large population around the world. In the United States (US) alone, every year, a significant percentage of the adult population is affected by different mental disorders, which include depression mental illness (6.7\%), anorexia and bulimia nervosa (1.6\%), and bipolar mental illness (2.6\%) \cite{american2013diagnostic}. Sometimes mental illness has been attributed to the mass shooting in the US \cite{metzl2015mental}, which has taken numerous innocent lives. One of the common mental health problems is depression that is more dominant than other mental illness conditions worldwide \cite{zafar2020survey}. The fatality risk of suicides in depressed people is 20 times higher than the general population \cite{WHO3}. Diagnosis of depression is usually a difficult task because depression detection needs a thorough and detailed psychological testing by experienced psychiatrists at an early stage \cite{rissola2020beyond} and it requires interviews, questionnaires, self-reports or testimony from friends and relatives. Moreover, it is very common among people who suffer from depression that they do not visit clinics to ask help from doctors in the early stages of the problem \cite{zou2020depression}. 

\begin{figure}[pos=htp]
\centering
\includegraphics[totalheight=5cm]{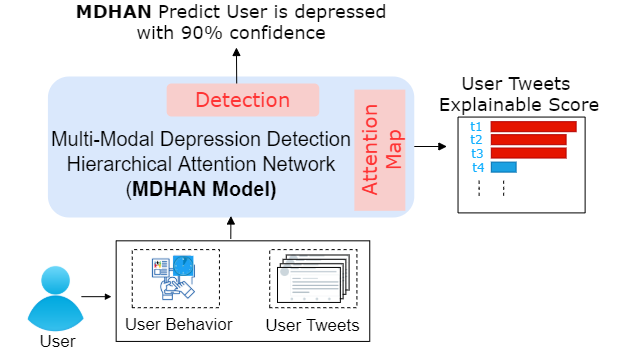}
\caption{Interpertable depression detection }
\label{Overview}
\end{figure}

Individuals and health organizations have thus shifted away from their traditional interactions, and now meeting online by building online communities for sharing information, seeking and giving the advice to help scale their approach to some extent so that they could cover more affected population in less time. Besides sharing their mood and actions, recent studies indicate that many people on social media tend to share or give advice on health-related information \cite{hawn2009take, neuhauser2003rethinking, scanfeld2010dissemination, prier2011identifying}. These sources provide the potential pathway to discover the mental health knowledge for tasks such as diagnosis, medications and claims. It is common for people who suffer from mental health problems to often ``implicitly'' (and sometimes even ``explicitly'') disclose their feelings and their daily struggles with mental health issues on social media as a way of relief \cite{park2012depressive, bathina2020depressed}. Therefore, social media is an excellent resource to automatically discover people who are under depression. While it would take a considerable amount of time to manually sift through individual social media posts and profiles to locate people going through depression, automatic scalable computational methods could provide timely and mass detection of depressed people which could help prevent many major fatalities in the future and help people who genuinely need it at the right moment.
\\

Usually, depressed users act differently when they are on social media, producing rich behavioural data, which is often used to extract various features. However, not all of them are related to depression characteristics. Recently deep learning has been applied for depression detection on social media and showed significantly better performance than traditional machine learning methods. Hamad et. al. presented a computational framework for automatic  detection of depressed user that initially selects relevant content through a hybrid extractive and abstractive summarization strategy on the sequence of all user tweets leading to a more fine-grained and relevant content, which then forwarded  to deep learning framework comprising of a unified learning machinery comprising of convolutional neural network coupled with attention-enhanced gated recurrent units leading to better empirical performance than existing strong baseline methods \cite{zogan2021depression}. Even though, recent work showed the effectiveness of deep learning methods for depression detection,  however, most of the existing machine learning methods provide no interpretability for depression prediction, hence their predictions are obscure to human which reduces the trust on the deep learning models. Interpretable model provides deep insight into how a deep learning model can be improved, and supports understanding. Thus, in order to engenders the appropriate user trust and provide the reason behind the decision, we aim develop a interpertable deep learning-based solution for depression detection by utilizing multi-modal features from diverse behaviour of the depressed user in social media. Apart from the latent features derived from lexical attributes, we notice that the dynamics of tweets, i.e. tweet timeline provides a crucial hint reflecting depressed user emotion change over time. To this end, we propose a hybrid model, Multi-Modal Depression Detection Hierarchical Attention Network \textbf{MDHAN} to boost the classification of depressed users using multi-modal features and word embedding features. Figure \ref{Overview} illustrate the effectiveness of interpertablity in improving the user trust. The model can derive new deterministic feature representations from training data and produce superior results for detecting depression-level of Twitter users,  and derive explanation from a user posts content. Beside, we also studied the performance of our model when we used the two components user posts and his multi-modalities features separately. We found that model performance deteriorated when we used only multi-modal features. We further show when we combined the two attributes, our model led to better performance. To summarize, our study makes the following \textbf{key contributions}:

\begin{enumerate}
  \item present a novel interpertable depression detection framework by deep learning the textual, behavioural, temporal, and semantic modalities from social media. To the best of our knowledge, this is the first work of using multi-modalities of topical, temporal and semantic features jointly with word embeddings in deep learning for depression detection. 
  
  \item  introduce the prospective of viewing explainablity of model for depression detection and build a pipeline aided with interpertablity  based on hierarchical attention networks to explain the prediction of depression detection. 

  \item Extensive experiment is performed on benchmark depression twitter dataset,  that shows superiority of our proposed method when compared to baseline methods.
\end{enumerate}

The rest of our paper is organized as follows. \textbf{Section 2} reviews the related work to our paper. \textbf{Section 3} presents the dataset that used in this work, and different pre-processing we applied on data. \textbf{Section 4} describes the two different attributes that we extracted for our model. In \textbf{Section 5}, we present our model for detection depression. \textbf{Section 6} reports experiments and results. Finally, \textbf{Section 7} concludes this paper.

\section{Related Work}
In this section, we will discuss closely related literature and mention how they are different from our proposed method. In general, just like our work, most existing studies focus on user behaviour to detect whether a user suffers from depression or any mental illness. We will also discuss other relevant literature covering word embeddings and hybrid deep learning methods which have been proposed for detecting mental health from online social networks and other resources including public discussion forums. Since we also introduce the notion of latent topics in our work, we have also covered relevant related literature covering topic modelling for depression detection, which has been widely studied in the literature.

Data present in social media is usually in the form of information that user shares for public consumption which also includes related metadata such as user location, language, age, among others \cite{kumar2019anxious}. In the existing literature, there are generally two steps to analyzing social data. The first step is collecting the data generated by users on networking sites, and the second step is to analyze the collected data using, for instance, a computational model or manually. In any data analysis, feature extraction is an important task because using only a relevant small set of features, one can learn a high-quality model.

Understanding depression on online social networks could be carried out using two complementary approaches which are widely discussed in the literature, and they are:
\begin{itemize}
    \item Post-level behavioural analysis
    \item User-level behavioural analysis
\end{itemize}


\subsection {Post-level behavioural analysis} Methods that use this kind of analysis mainly target at the textual features of the user post that is extracted in the form of statistical knowledge such as those based on count-based methods \cite{lebret2015rehabilitation}. These features describe the linguistic content of the post which are discussed in \cite{de2013predicting, hu2015predicting}. For instance, in \cite{de2013predicting} the authors propose classifier to understand the risk of depression. Concretely, the goal of the paper is to estimate that there is a risk of user depression from their social media posts. To this end, the authors collect data from social media for a year preceding the onset of depression from user-profiles and distil behavioural attributes to be measured relating to social engagement, emotion, language and linguistic styles, ego network, and mentions of antidepressant medications. The authors collect their data using crowd-sourcing task, which is not a scalable strategy, on Amazon Mechanical Turk. In their study, the crowd workers were asked to undertake a standardized clinical depression survey, followed by various questions on their depression history and demographics. While the authors have conducted thorough quantitative and qualitative studies, they are disadvantageous in that it does not scale to a large set of users and does not consider the notion of text-level semantics such as latent topics and semantic analysis using word embeddings. Our work is both scalable and considers various features which are jointly trained using a novel hybrid deep learning model using a multi-modal learning approach. It harnesses high-performance Graphics Processing Units (GPUs) and as a result, has the potential to scale to large sets of instances. In Hu et al., \cite{hu2015predicting} the authors also consider various linguistic and behavioural features on data obtained from social media. Their underlying model relies on both classification and regression techniques for predicting depression while our method performs classification, but on a large-scale using a varied set of crucial features relevant to this task.

To analyze whether the post contains positive or negative words and/or emotions, or the degree of adverbs \cite{tsugawa2015recognizing} used cues from the text, for example, \textit{I feel a little depressed} and \textit{I feel so depressed}, where they capture the usage of the word \textit{``depressed''} in the sentences that express two different feelings. The authors also analyzed the posts' interaction (i.e., on Twitter (retweet, liked, commented)). Some researchers studied post-level behaviours to predict mental problems by analysing tweets on Twitter to find out the depression-related language. In \cite{resnik-etal-2015-beyond}, the authors have developed a model to uncover meaningful and useful latent structure in a tweet. Similarly, in \cite{shen2017depression}, the authors monitored different symptoms of depression that are mentioned in a user's tweet. In \cite{soton423226}, they study users' behaviour on both Twitter and Weibo. To analyze users' posts, they have used linguistic features. They used a Chinese language psychological analysis system called TextMind in sentiment analysis. One of the interesting post-level behavioural studies was done by \cite{shen2017depression} on Twitter by finding depression relevant words, antidepressant, and depression symptoms. In \cite{ramirez2018early} the authors used post-level behaviour for detecting anorexia; they analyze domain-related vocabulary such as anorexia, eating disorder, food, meals and exercises.

\subsection{User-level behaviours} There are various features to model users in social media as it reflects overall behaviour over several posts. Different from post-level features extracted from a single post, user-level features extract from several tweets during different times \cite{tsugawa2015recognizing}. It also extracts the user's social engagement presented on Twitter from many tweets, retweets and/or user interactions with others. Generally, posts' linguistic style could be considered to extract features \cite{hu2015predicting, yazdavar2017semi, yazdavar2017semi}. The authors in \cite{shen2017depression} extracted six depression-oriented feature groups for a comprehensive description of each user from the collected data set. The authors used the number of tweets and social interaction as social network features. For user profile features, they have used user shared personal information in a social network. Analysing user behaviour looks useful for detecting eating disorder. In Wang et al., \cite{wang2017detecting} they extracted user engagement and activities features on social media. They have extracted linguistic features of the users for psychometric properties which resembles the settings described in \cite{ramirez2018early, kumar2019anxious, soton423226} where the authors have extracted 70 features from two different social networks (Twitter and Weibo). They extracted features from a user profile, posting time and user interaction feature such as several followers and followee. Similarly, Wong et al.  combined user-level and post-level semantics and cast their problem as a multiple instance learning setup. The advantage that this method has is that it can learn from user-level labels to identify post-level labels \cite{wongkoblap2019modeling}.


 Recently, Lin et al. applied CNN-based deep learning model to classify Twitter users based on depression using multi-modal features. The framework proposed by the authors has two parts. In the first part, the authors train their model in an offline mode where they exploit features from Bidirectional Encoder Representations from Transformers (BERT) \cite{devlin2018bert} and visual features from images using a CNN model. The two features are then combined, just as in our model, for joint feature learning. There is then an online depression detection phase that considers user tweets and images jointly where there is a feature fusion at a later stage. In another recently proposed work \cite{chiu2020multimodal}, the authors use visual and textual features to detect depressed users on Instagram posts than Twitter. Their model also uses multi-modalities in data, but keep themselves confined to Instagram only. While the model in \cite{lin2020sensemood} showed promising results, it still has certain disadvantage. For instance, BERT vectors for masked tokens are computationally demanding to obtain even during the fine-tuning stage, unlike our model which does not have to train the word embeddings from scratch. Another limitation of their work is that they obtain sentence representations from BERT, for instance, BERT imposes a 512 token length limit where longer sequences are simply truncated resulting in some information loss, where our model has a much longer sequence length which we can tune easily because our model is computationally cheaper to train. We have proposed a hybrid model that considers a variety of features unlike these works. While we have not specifically used visual features in our work, using a diverse set of crucial relevant textual features is indeed reasonable than just visual features. Of course, our model has the flexibility to incorporate a variety of other features including visual features.

Multi-modal features from the text, audio, images have also been used in \cite{zheng2020graph}, where a new graph attention-based model embedded with multi-modal knowledge for depression detection. While they have used temporal CNN model, their overall architecture has experimented on small-scale questionnaire data. For instance, their dataset contains 189 sessions of interactions ranging between 7-33min (with an average of 16 min). While they have not experimented their method with short and noisy data from social media, it remains to be seen how their method scales to such large collections. Xezonaki et al., \cite{xezonaki2020affective} propose an attention-based model for detecting depression from transcribed clinical interviews than from online social networks. Their main conclusion was that individuals diagnosed with depression use affective language to a greater extent than those who are not going through depression. In another recent work \cite{wolohan2020estimating}, the authors discuss depression among users during the COVID-19 pandemic using LSTM and fastText \cite{mikolov2017advances} embeddings. In \cite{shrestha2020multi}, the authors also propose a multi-model RNN-based model for depression prediction but apply their model on online user forum datasets. Trotzek et al., \cite{trotzek2018utilizing} study the problem of early detection of depression from social media using deep learning where the leverage different word embeddings in an ensemble-based learning setup. The authors even train a new word embedding on their dataset to obtain task-specific embeddings. While the authors have used the CNN model to learn high-quality features, their method does not consider temporal dynamics coupled with latent topics, which we show to play a crucial role in overall quantitative performance. Farruque et al., \cite{farruque2019augmenting} study the problem of creating word embeddings in cases where the data is scarce, for instance, depressive language detection from user tweets. The underlying motivation of their work is to simulate a retrofitting-based word embedding approach \cite{faruqui2014retrofitting} where they begin with a pre-trained model and fine-tune the model on domain-specific data.

Topic modeling has also been used for detection of depression from social media. Gong et al., \cite{gong2017topic} proposed a topic modelling approach to depression detection using multi-modal analysis. They propose a novel topic model which is context-aware with temporal features. While the model produced satisfactory results on 2017 Audio/Visual Emotion Challenge (AVEC), the method does not use a variety of rich features and could face scalability issues because simple posterior inference algorithms such as those based on Gibbs or collapsed Gibbs sampling do not parallelize unlike deep learning methods, or one need sophisticated engineering to parallelize such models.


Recent studies have started to target depressed user online, extracting features representing user behaviours and classifying these features into different groups, such as the number of posts, posting time distribution, and number followers and followee. Peng et. al. extracted different features and classified them into three groups, user profile, user behaviour and user text and used multi-kernel SVM for classification \cite{peng2019multi}. The above-mentioned works have some limitations. They mainly focused on studying user behaviour than taking cues from user generated content such as the text they share which make it extremely difficult to achieve high performance in classification. These models also cannot work well to detect depressed user at user-level, and as a result, they are prone to incorrect prediction. Our novel approach combines user behaviour with user history posts. Besides, our strategy to select salient content using automatic summarization helps our model only focus on the most important information.
Although, recent deep learning methods showed significant performance for depression detection,  however, most of the existing machine learning methods provide no interpretability for prediction. Interpretable model can provide deep insight into how a deep learning model can be improved, and supports understanding. Therefore, to provide some details and explain user tweets or reasons to make decision functioning clear or easy to understand, we aim to develop an interpretable deep learning-based approach for depression detection. Our proposed model utilized multi-modal features from the diverse behaviour of the depressed user and his posts in social media.

\section{Problem Statement}
To formulate our problem, suppose that we have a set $U$ of labeled users either depression or non-depression samples. Let A be a user posts $A=[t_1, t_2,....,t_L]$ consisting L tweets, where L is the total number of tweets per user, each tweet $t_i$ contains n words $t_i=[w_{i1}, w_{i2},....,w_{iN}]$ where N is the total number of words per tweet. And Let M be the features in total for a user \{$m_i$\}$^M _{i=1}$, and let \{$1,2,...,S$\} be a  finite set of available modalities, so we donate $M_s$ as the dimension of $S^{th}$ modality. Therefore, once we have a user tweets A and a set of related a user behaviours features M, Learn depression detection function will be as following:
 \begin{equation}
f (A,M) \rightarrow \hat{y}
\end{equation}

such that it maximizes prediction accuracy. And in our problem  we treat depression detection as the binary classification problem, i.e., user can be depressed  ($\hat{y}$ = 1) or not-depressed  ($\hat{y}$ = 0).

\section{Dataset} \label{data_twitter}


Developing and validating the terms used in the vocabulary by users with mental illness is time-consuming but helps obtain a reliable list of words, by which reliable tweets could be crawled reducing the amount the false-positives. Recent research conducted by the authors of \cite{shen2017depression} is one such work that has collected a large-scale data with reliable ground truth data, which we aim to reuse. We present the statistics of the data in \textit{\textbf{Table}}~\ref{table1}. To exemplify the dataset further, the authors collected three complementary data sets, which are:
\begin{itemize}
    \item Depression data set: Each user is labelled as depressed, based on their tweet content between 2009 and 2016. This includes 1,402 depressed users and 292,564 tweets.
    \item Non-depression data set: Each user is labelled as non-depressed and the tweets were collected in December 2016. This includes over 300 million active users and 10 billion tweets.
    \newpage
    \item Depression-candidate data set: The authors collected are labelled as depression-candidate, where the tweet was collected if contained the word ``depress''. This includes 36,993 depression-candidate users and over 35 million tweets.
\end{itemize}

\begin{table}
\centering
\begin{tabular}{l l l}
\hline
\textbf{Dataset } & \textbf{Depressed} & \textbf{Non-Depressed}\\
\hline
No. of Users  & 1402 & 300 million \\
No. of Tweets & 292,564 & 10 billion \\
\hline
\end{tabular}
\caption{Statistics of the large dataset collected by the authors in \cite{shen2017depression} which is used in this study.}
\label{table1}
\end{table}

Data collection mechanisms are often loosely controlled, impossible data combinations, for instance, users labelled as depressed but have provided no posts, missing values, among others. After data has been crawled, it is still not ready to be used directly by the machine learning model due to various noise still present in data, which is called the ``raw data''. The problem is even more exacerbated when data has been downloaded from online social media such as Twitter because tweets may contain spelling and grammar mistakes, smileys, and other undesirable characters. Therefore, a pre-processing strategy is needed to ensure satisfactory data quality for computational modal to achieve reliable predictive analysis.



To further clean the data we used Natural Language processing ToolKit (NLTK). This package has been widely used for text pre-processing \cite{article} and various other works. It has also been widely used for removing common words such as stop words from text \cite{8389299, kumar2019anxious, resnik-etal-2015-beyond}. We have removed the common words from users tweets (such as ``the'', ``an'', etc.) as these are not discriminative or useful enough for our model. These common words sometimes also increase the dimensionality of the problem which could sometimes lead to the ``curse-of-dimensionality'' problem and may have an impact on the overall model efficiency. To further improve the text quality, we have also removed non-ASCII characters which have also been widely used in literature \cite{yazdavar2017semi}.

Pre-processing and removal of noisy content from the data helped get rid of plenty of noisy content from the dataset. We then obtained a high-quality reliable data which we could use in this study. Besides, this distillation helped reduce the computational complexity of the model because we are only dealing with informative data which eventually would be used in modelling. We present the statistics of this distilled data below:

\begin{itemize}
\item Number of users labelled positive tweets: 5899.
\item Number of tweets from positive users: 508786.
\item Number of users labelled negative: 5160.
\item Number of tweets from negative users: 2299106.
\end{itemize}

To further mitigate the issue of sparsity in data, we excluded those users who have posted less than ten posts and users who have less than 5000 followers, therefore we ended up with 2500 positive users and 2300 negative users. 

\section{Explainable Deep Depression Detection}

\begin{figure*}
\centering
\includegraphics[totalheight=6.5cm]{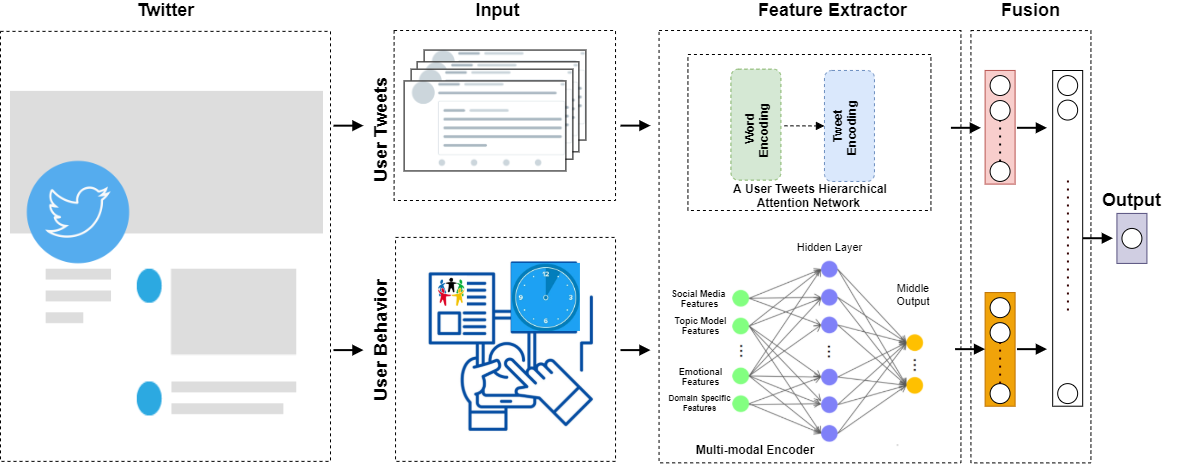}
\caption{Overview of our proposed model MDHAN: We predict depressed user by fusing two kinds of information: (1) User tweets. (2) User Behaviours. }
\label{MDHAN}
\end{figure*}
Due to the complexity of user posts and the diversity of their behaviour on social media, we propose a hybrid model based on Hierarchical Attention Networks that combines with Multilayer perceptron MLP to detect depression through social media as depicted in Figure~\ref{MDHAN}. For each user, the model takes two inputs for the two attributes. First, the four modalities feature input that represents user behaviour vector runs into MLP, capturing distinct and latent features and correlation across the features matrix. The second input represents each user input tweets that will be replaced with it's embedding and fed to Hierarchical Attention Networks to learn some representation features through a hierarchical word and tweet level encoding. The output in the middle of both attributes is concatenated to represent one single vector feature that fed into an activation layer of sigmoid for prediction. In the following sections, we will discuss the following two existing separate architectures.
\subsection{Features Selection}
We introduce this attribute type where the goal is to calculate the attribute value corresponding to each modality for each user. We estimate that the dimensionality for all modalities of interest is 76; and we mainly consider four major modalities as listed below and ignore two modalities due to missing values. These features are extracted respectively for each user as follows:

\subsubsection {\textbf{Social Information and Interaction:}}
From this attribute, we extracted several features embedded in each user profile. These are features related to each user account as specified by each feature name. Most of the features are directly available in the user data, such as the number of users following and friends, favourites, etc.

Moreover, the extracted features relate to user behaviour on their profile. For each user, we calculate their total number of tweets, their total length of all tweets and the number retweets. We further calculate posting time distribution for each user, by counting how many tweets the user published during each of the 24 hours a day. Hence it is a 24-dimensional integer array. To get posting time distribution for each tweet, we extract two digits as hour information, then go through all tweets of each user and track the count of tweets posted in each hour of the day.

\subsubsection{ \textbf{Emojis Sentiment:}}
Emojis allow users to express their emotions through simple icons and non-verbal elements. It is useful to get the attention of the reader. Emojis could give us a glance for the sentiment of any text or tweets, and it is essential to differentiate between positive and negative sentiment text \cite{novak2015sentiment}. User tweets contain a large number of emojis which can be classified into positive, negative and neutral. For each positive, neutral, and negative type, we count their frequency in each tweet. Then we sum up the numbers from each user's tweets to get the sum for each user. So the final output is three values corresponding to positive, neutral and negative emojis by the user. We also consider Voice Activity Detection (VAD) features. These features contain Valance, Arousal and Dominance scores. For that, we count First Person Singular and First Person Plural. Using affective norms for English words, a VAD score for 1030 words are obtained. We create a dictionary with each word as a key and a tuple of its (valance, arousal, dominance) score as value. Next, we parse each tweet and calculate VAD score for each tweet using this dictionary. Finally, for each user, we add up the VAD scores of tweets by that user, to calculate the VAD score for each user.

\subsubsection{ \textbf{Topic Distribution:}}
Topic modelling belongs to the class statistical modelling frameworks which helps in the discovery of abstract topics in a collection of text documents. It gives us a way of organizing, understanding and summarizing collections of textual information. It helps find hidden topical patterns throughout the process, where the number of topics is specific by the user apriori. It can be defined as a method of finding a group of words (i.e. topics) from a collection of documents that best represent the latent topical information in the collection. In our work, we applied the unsupervised Latent Dirichlet Allocation (LDA) \cite{blei2003latent} to extract the most latent topic distribution from user tweets. To calculate topic level features, we first consider corpus of all tweets of all depressed users. Next, we split each tweet into a list of words and assemble all words in decreasing order of their frequency of occurrence, and common English words (stopwords) are removed from the list. Finally, we apply LDA to extract the latent \(K=25\) topics distribution, where \(K\) is the number of topics. We have found experimentally \(K=25\) to be a suitable value. While there are tuning strategies and strategies based on Bayesian non-parametrics \cite{teh2005sharing}, we have opted to use a simple, popular, and computationally efficient approach which helps give us the desired results.

\subsubsection{ \textbf{Symptoms features}} 1- Depression symptom counts: It is the count of depression symptoms occurring in tweets, as specified in nine groups in DSM-IV criteria for a depression diagnosis. The symptoms are listed in Appendix \ref{appendix:symptoms}. We count how many times the nine depression symptoms are mentioned by the user in their tweets. The symptoms are specified as a list of nine categories, each containing various synonyms for the particular symptom. We created a set of seed keywords for all these nine categories, and with the help of the pre-trained word embedding, we extracted the similarities of these symptoms to extend the list of keywords for each depression symptoms. Furthermore, we scan through all tweets, counting how many times a particular symptom is mentioned in each tweet. 2- Antidepressants: We also focused on the antidepressants, and we created a lexicon of antidepressants from the ``Antidepressant'' Wikipedia page which contains an exhaustive list of items and is updated regularly, in which we counted the number of names listed for antidepressants. The medicine names are listed in Appendix \ref{appendix:antidepressant}.

\subsection{User Tweets Encoder Using RNN}
\begin{figure*}[pos=htp]
\centering
\includegraphics[totalheight=3.4cm]{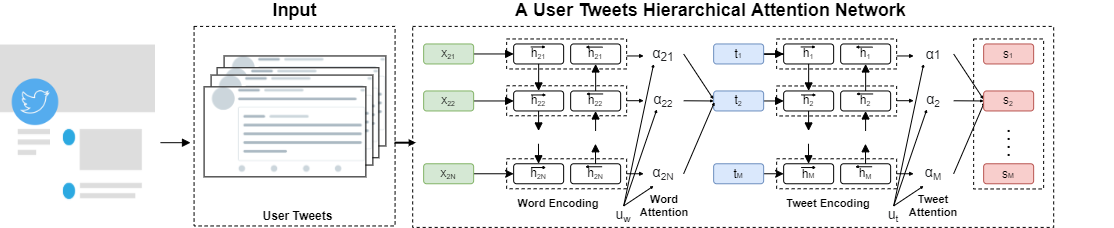}
\caption{An illustration of hierarchical attention network that we used to encode user tweets}
\label{HAN}
\end{figure*}

Recently, researchers find that the Hierarchical Attention Network (HAN)\cite{yang2016hierarchical} is useful and able to find the most important words and sentences in a document while considering it. A depressed user could often have different linguistic style posts, including depressive language use, and mentions of antidepressants and symptoms, which can help detect depression. 
Additionally, a social media post contains linguistic prompts with different levels of word-level and tweet-level. Every word in a tweet and every tweet of a user is equally important to understand a depressed user in social media. For example, "My dad doesn't even seem to believe I'm really hurt!", the word "hurt" contributes more signals to decide whether the tweet is depressed rather than other words in the tweet. So in this way, HAN performs better in predicting the class of given user tweets. Inspired by \cite{yang2016hierarchical}, we proposed Hierarchical Attention Network to learn user tweets representation as depicted in Figure \ref{HAN}. We consider U be a user made M tweets $T=[t_1, t_2,....,T_M]$
each tweet $t_i=[w_1, w_2,....,w_N]$ contains $N_i$ words. Each tweet is represented by the sequence of d-dimensional embeddings of their words, $t_i=[w_{11},....,w_{MN}]$. And we represent each word as the input layer a fixed-size vector from pre-trained word embeddings.
\\
\\
\subsubsection{ \textbf{Word Encoder:}} A bidirectional Gated Recurrent Unit is first used as the word level encoder to capture annotations' contextual information. GRU is a Recurrent Nural Network (RNN) that can capture sequential information and sentences' long-term dependency. Only two gate functions are used which are reset and update gates. Update gate has been used to monitor the degree with which the previous moment's status information has transported into the current state. The higher the update gate value, the more the previous moment's status information is carried forward. The reset gate has been used to monitor the degree with which the previous moment's status information is overlooked. The smaller the reset gate value, the more neglected the context will be. Both the preceding and the following words influence the current word in the sequential textual data, so we use the BiGRU model to extract the contextual features. The BiGRU consists of a forward $\overrightarrow{GRU}$ and a backward $\overleftarrow{GRU}$ that are used, respectively, to process forward and backward data. The annotation $w_{ij}$ represent the word j in a sentence i that contains N-words. Each word of user post (tweet) will convert to a word embedding $x_{ij}$ utilising GloVe \cite{pennington2014glove}.

 \begin{equation}
  {\overrightarrow{ h_{ij}^w}} = \overrightarrow{GRU}\left( {{x_{ij}}} ,\overrightarrow{ h_{i(j-1)}}\right),  j \in \{1,..., N\}\end{equation}

   \begin{equation}
  {\overleftarrow{ h_{ij}^w}} = \overleftarrow{GRU}\left( {{x_{ij}}} ,\overleftarrow{ h_{i(j-1)}}\right),  j \in \{N,..., 1\}\end{equation}
  \\
  The combination of the hidden state that is obtained from the forward GRU and the backward GRU \(\overrightarrow {h_{ij}^w} \) and \( \overleftarrow {h_{ij}^w}\) is represented as \( h_{ij}^w \) $=\left[{\overrightarrow{ h_{ij}^w}} \oplus {\overleftarrow{ h_{ij}^w}}\right]$. Which carries the complete tweet information centred around ${x_{ij}}$.
\\
\\
Next, we describe the attention mechanism. It is crucial to introduce a vector $u_{ij}$ for all words, which is trainable and expected to capture global word. The \( {h_{ij}^w} \) annotations create the basis for attention that starts with another hidden layer by letting the model learn and randomly initialized biases and weights through training. the annotations $u_{ij}$ will be represented as following:
   \begin{equation}
 u_{ij} = tanh (W_w h_{ij}^w + b_w)\end{equation}
 The product ${u_{ij} u_w}$ (${u_w}$ is randomly initialized) expected to signal the importance of the $j$ word and normalized to an importance weight per word $\alpha_{ij}$ by a softmax function:
  \begin{equation}
\alpha_{ij}= \frac{\exp(u_{ij} u_w)}{\sum\limits_{j}\exp({ij} u_w)}
\end{equation}
Finally, a weighted sum of word representations concatenated with the annotations previously determined called the tweet vector $v_i$, where $\alpha_{t}$ indicating importance weight per word: 
\begin{equation}
v_{i}= \sum\nolimits_{t} \alpha_{ij} h_{ij}^w
\end{equation}

\subsubsection{ \textbf{Tweet Encoder:}} In order to learn the tweet representations $h_i^t$ from a learned tweet vector $v_i$, we capture the information of context  at the tweet level. Similar to the  word encoder component, tweet encoder employs the same BiGRU architecture. Hence the combination of the hidden state that is obtained from the forward GRU and the backward GRU \(\overrightarrow {h_{i}^t} \) and \( \overleftarrow {h_{i}^t}\) is represented as \( h_{i}^t \) $=\left[{\overrightarrow{ h_{i}^t}} \oplus {\overleftarrow{ h_{i}^t}}\right]$. Which capture the coherence of a tweet concerning its neighbouring tweets in both directions. Next, we will capture the related tweets in the form vector $\hat{t}$ by using tweet level attention layer. The product ${u_{i} u_s}$  expected to signal the importance of the $i$ tweet and normalized to an importance weight per tweet $\alpha_{i}$. Finally, $s_i$ will be a vector that summarizes all the tweet information in a user posts: 
\begin{equation}
s_{i}= \sum\nolimits_{t} \alpha_{i} h_{i}^t
\end{equation}







\subsection{Multi-modal Encoder:} 

Suppose the input which resembles a user behaviour be represented as [\(m_{1}, m_{2}, . . . ,  m_{M} \) ] where M is the total number of features and $M_s$ is the dimension of $S^{th}$ modality. Hence, to obtain fine-grained information from user behaviours features, the multi-modal features are fed through a one-layer MLP to get a hidden representation  $m_i$:

\begin{equation} p_{i}=f(b + \sum\limits_{i=1}^{M} W_i m_i ) \end{equation}

 Where f stands for the nonlinear function and the outcome of behaviour modelling $p_i$ is the high-level representation that captures the behavioural semantic information and plays critical role in depression diagnosis.

\subsection{Classification Layer:} 
At the classification layer, we need to
predict whether the user is depressed or not depressed. So far, we have introduced, how we encode user multi\_modalities behaviors (${p}$) and how we can encode user tweets by modeling the hierarchical structure from word level and tweet level (${s}$). Then From both component we construct the feature matrix of user behaviours features and user tweets:

\begin{equation} 
p = \{p_1,p_2, .... , p_M\} \in {\mathbb{R}^{1d \times M}}
\end{equation}

\begin{equation} 
s = \{s_1,s_2, .... , s_n\} \in {\mathbb{R}^{2d \times n}}   \end{equation}

We further integrate these components together $[{p},{s}]$. And The output of such a network is typically fed to a sigmoid layer for classification:
 \begin{equation}
\hat{y}= Sigmoid (b_f+[{p},{s}]W_f)
\end{equation}
Where where $\hat{y}$ is the predicted probability vector with $\hat{y_0}$ and $\hat{y_1}$ indicate the predicted probability of label being 0 (not depressed)
and 1 (depressed user) respectively. Then, we aim to minimize the cross-entropy error for each user with ground-truth label $y$ :
\[\mathrm{Loss} = - \sum_{i} y_i \cdot \mathrm{log}\; {\hat{y}}_i \]
where \(\hat{y}_i\) is the predicted probability and \(y_i\) is the ground truth label (either depression or non-depression) user.

\section{Experiments and Results}
In this section, we present the experimental evaluation to validate the performance  of \textbf{MDHAN}. First will we will introduce Datasets and Evaluation Metrics and Experimental Setting, followed by the Experimental Results. We compare our model with the following classification methods:


\begin{itemize}
\item \textbf{MDL: Multimodal Dictionary Learning Model} is to detect depressed users on Twitter \cite{shen2017depression}. They use a dictionary learning to extract latent data features and sparse representation of a user.
\item \textbf{SVM: Support Vector Machine} is a popular and a strong classifier that has been applied on a wide range of classification tasks \cite{karmen2015screening} and it still remains a strong baseline. 

\item \textbf{NB: Naive Bayes} is a family of probabilistic algorithms based on applying Bayes' theorem with the ``naive'' assumption of conditional independence between instances \cite{ng2002discriminative}.

\item \textbf{BiGRU}:We applied \textbf{Bidirectional Gated Recurrent Unit} \cite{cho-etal-2014-learning} with attention mechanism to obtain user tweets representations, which we then used for user tweets classification.

\item \textbf{MBiGRU}: Hybrid model based on MLP and BiGRU for Multi-Modal features for the user behaviour and the user’s online timeline (posts).

\item \textbf{CNN}: We utilized \textbf{Convolutional Neural Networks} \cite{kim-2014-convolutional} with an attention mechanism to model user tweets, which can capture the semantics of different convolutional window sizes for depression detection.

\item \textbf{MCNN}: Hybrid model based on MLP and CNN for Multi-Modal features for the user behaviour and the user’s online timeline (posts).

\item \textbf{HAN}: A hierarchical attention neural network framework \cite{yang2016hierarchical}, it used on user posts for depression detection. The network encodes first user posts with word level attentions on each tweet and tweet-level attentions on each user posts.
\item \textbf{MDHAN}: The proposed model in this paper.

\end{itemize}

\subsection{Datasets and Evaluation Metrics}
For our experiments, we have used the datasets as mentioned in section (3). They provide a large scale of data, especially for labelled negative and candidate positive, and in our experiments, we used the labelled dataset. We preprocess the dataset by excluding users who have their posting history comprising of less than ten posts or users with followers more than 5000, or users who tweeted in other than English so that we have sufficient statistical information associated with every user. We have thus considered 4208 users (51.30\% depressed and 48.69 \% non-depressed users) as shown in Table \ref{tab:data}. For evaluation purpose, we split the dataset into training (80\%) and test (20\%). 

\begin{table} [pos=htbp]
  \centering
    \begin{tabular}{l|c|c}
    \toprule
    \toprule
    Description & Depressed & Non-Depressed \\
    \midrule
    Numer of users & 2159  & 2049 \\
    Number of tweets & 447856 & 1349447 \\
    \bottomrule
    \bottomrule
    \end{tabular}%
        \caption{Summary of labelled data used to train MDHAN model} 
  \label{tab:data}%
\end{table}%

Additionally, We employ traditional information retrieval metrics such as precision, recall, F1, and accuracy based on the confusion matrix to evaluate our model. A confusion matrix is a sensational matrix used for evaluating classification performance, which is also called an error matrix because it shows the number of wrong predictions versus the number of right predictions in a tabulated manner. 

\subsection{Experimental Setting}
For parameter configurations, the word embeddings are initialized with the Glove \cite{pennington2014glove} with a dimension of 100 on the training set of each dataset to initialize the word embeddings of all the models, including baselines. The hidden dimension has set to 100 in our
model and other neural models, also, the dropout is set to 0.5. All the models are trained to use use the Adam optimization algorithm \cite{Kingma2015AdamAM} with a batch size of 16 and an initial learning rate an initial learning rate of 0.001. Finally we trained our model for 10 iterations, with batch size of 16. The number of iterations was sufficient to converge the model and our experimental results further cement this claim where we outperform existing strong baseline methods, and the training epoch is set to
10. We used python 3.6.3 and Tensorflow 2.1.0 to develop our implementation. We rendered the embedding layer to be not trainable so that we keep the features representations, e.g., word vectors and topic vectors in their original form. We used one hidden layer, and max-pooling layer of size 4 which gave better performance in our setting.

\begin{table*}[]
\begin{tabular}{lccccccccc}
\hline
Matric    & SVM   & NB    & MDL   & BiGRU & MBiGRU & CNN   & MCNN  & HAN   & \textbf{MDHAN} \\ \hline
Accuracy  & 0.644 & 0.636 & 0.787 & 0.764 & 0.786  & 0.806 & 0.871 & 0.844 & \textbf{0.895} \\
Precision & 0.724 & 0.724 & 0.790 & 0.766 & 0.789  & 0.817 & 0.874 & 0.870 & \textbf{0.902} \\
Recall    & 0.632 & 0.623 & 0.786 & 0.762 & 0.787  & 0.804 & 0.870 & 0.840 & \textbf{0.892} \\
F1-score        & 0.602 & 0.588 & 0.786 & 0.763 & 0.786  & 0.803 & 0.870 & 0.839 & \textbf{0.893} \\ \hline
\end{tabular}
\caption{Performance comparison of MDHAN against the baselines for depression detection on \cite{shen2017depression} Dataset }
\label{table performance}
\end{table*}

\begin{figure*}
  \centering
  
  \begin{minipage}[b]{0.45\textwidth}
    \includegraphics[width=\textwidth]{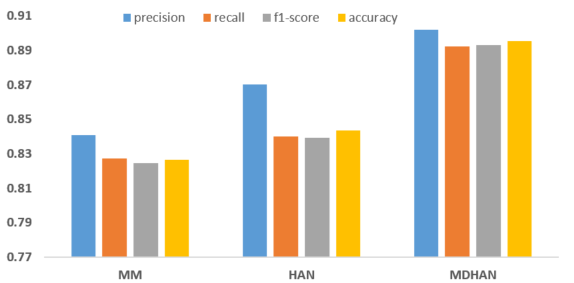}
     \caption{ Effectiveness comparison between MDHAN\\ with different attributes.}
    \label{figure different attributes}
  \end{minipage}
  \hfill
  \begin{minipage}[b]{0.45\textwidth}
    \includegraphics[width=\textwidth]{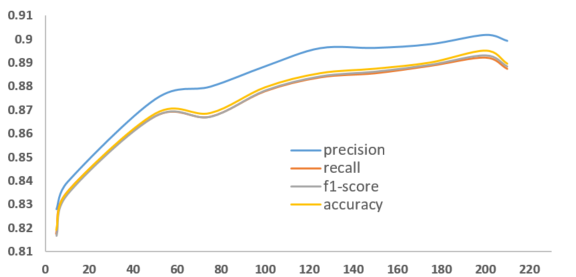}
    \caption{Model vs number of tweets}
    \label{number of tweets}
  \end{minipage}
    \begin{minipage}[b]{0.45\textwidth}
    \includegraphics[width=\textwidth]{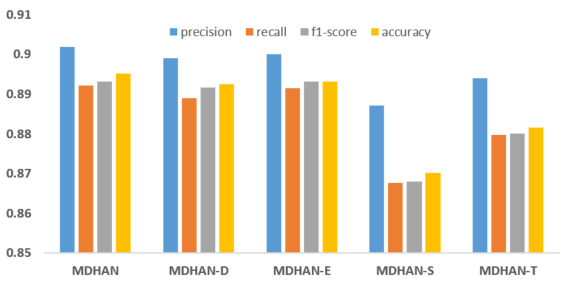}
    \caption{Modality effectiveness}
    \label{minus}
  \end{minipage}
    \hfill
    \begin{minipage}[b]{0.45\textwidth}

    \includegraphics[width=\textwidth]{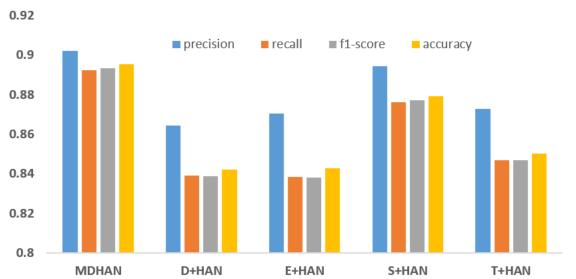}
     \caption{Modality effectiveness}
    \label{plus}

  \end{minipage}
\end{figure*}



\subsection{Experimental Results}
In our experiments, we study our model attributes including the quantitative performance of our hybrid model. The multi-modal features and user's timeline semantic features attribute, we will use both these attributes jointly. After grouped user behaviour in social media into multi-modal attribute, we evaluate the performance of the model. First, we examine the effectiveness of using the multi-Modal features only for Depression Detection with different classifiers. Second, we showed how the model performance increased when we utilize multi-modal features with hierarchical attention network MDHAN. We summarise the results in Table \ref{table performance} as follows:

\begin{itemize}
    \item Naive Bayes obtain the lowest F1 score, which demonstrates that this model has less capability to classify tweets when compared with other existing models to detect depression. The reason for its poor performance could be that the model is not robust enough to sparse and noisy data.
    \item MDL model outperforms SVM, NB and BiGRU, and obtains better accuracy than these three methods. Since this is a recent model especially designed to discover depressed users, it has captured the intricacies of the dataset well and learned its parameters faithfully leading to better results.

    \item we can observe the evolving when we integrate The multi-modal features with user posts and that better helped to analyze a user that seems to be depressed as shown in the performance of MBiGRU, MCNN MDHAN. 
    
    \item We can see our proposed model MDHAN improved the depression detection up to 10\% on F1-Score, compared to MDL model and 5\% compared to HAN model. This suggests that our model outperforms a strong model. The reason why our model performs well is primarily because it leverages a rich set of features which is jointly learned in the consolidated parameters estimation resulting in a robust model.

    \item Furthermore, MDHAN achieved the best performance with 89\% in F1, indicating that combining HAN with multimodal strategy for user timeline semantic features strategy is sufficient to detect depression in Twitter. We can also deduce from the table that our model consistently outperforms all existing and strong baselines.

\end{itemize}

\subsection{Comparison and Discussion}


To get a better look for our model performance, We have  compared the effectiveness of each of the two attributes of our model. Therefore, we test the performance of the model with a different attribute, we build the model to feed it with each attribute separately and compare how the model performs. First, we test the model using only the multi-modalities attribute, we can observe in Fig~\ref{figure different attributes} the model perform less optimally when we used MLP for multi-modal features. In contrast, the model performs better when we use only HAN with word embedding attribute. This signifies that extracting semantic information features from user tweets is crucial for depression detection. Thus, we can see the MDHAN model performance increased when combined both MM and HAN, and outperforms when using each attribute independently. One of the key parameters in MDHAN is the number of tweets for each user; we eventually observed that MDHAN reached optimal performance when using 200 tweets as the maximum number of tweets. Figure \ref{number of tweets} illustrates the performance of our model concerning the number of tweets.

To further analyze the role played by each modal features and contribution of the user behavioural attributes and user posts attribute, we removed the four modalities separately as following: the domain-specific feature and denote as \emph{MDHAN - D}, emotion feature and denote as \emph{MDHAN - E}, the social network feature and denote this model as \emph{MDHAN - S} and topic feature which we denote as \emph{MDHAN - T}. We can see in Figure \ref{minus} that our model performance deteriorates as we remove the topic feature from MDHAN model and degrades more without the social network features. To dive deeper and understand the effectiveness of each modal, we combine each modal separately with HAN and denote them respectively as following: \emph{D+HAN}, \emph{E+HAN}, \emph{S+HAN} and \emph{T+HAN}. As shown in Figure \ref{plus}, we could see that MDHAN with four Modalities outperforms the others, which means that each modality do contribute to depression detection.

\begin{figure*}[pos=h]
\centering
\includegraphics[totalheight=3.4cm]{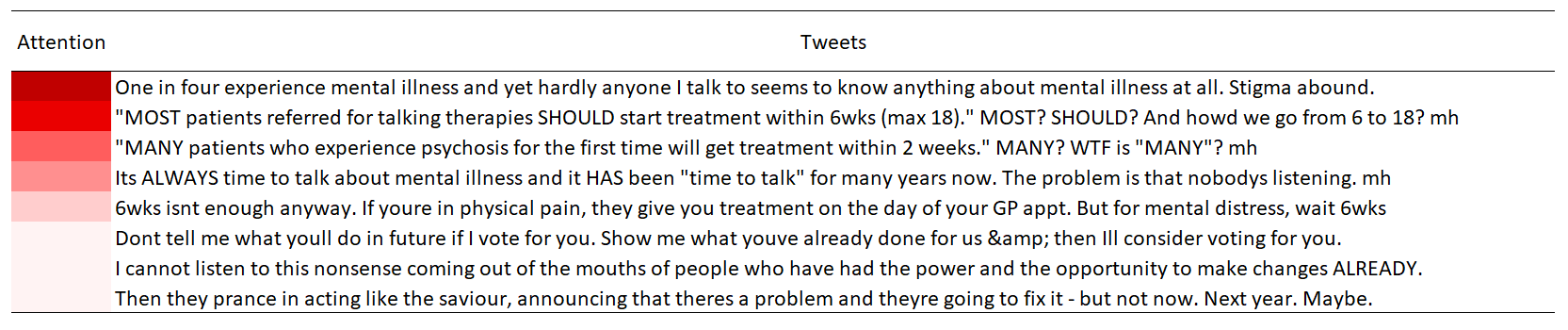}
\caption{Explainable via visualization of attention score in MDHAN}
\label{attention_score}
\end{figure*}

\begin{figure}%
    \centering
    \subfloat[\centering A word cloud for Suicidal ideation. ]{{\includegraphics[width=4.3cm]{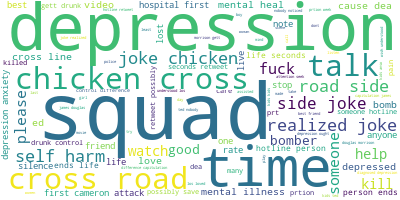} }}%
    \subfloat[\centering A word cloud for Worthlessness. ]{{\includegraphics[width=4.3cm]{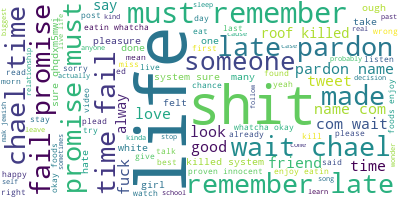} }}%
    
    \vspace{0.1cm}  
    \subfloat[\centering A word cloud for Loss of Energy. ]{{\includegraphics[width=4.3cm]{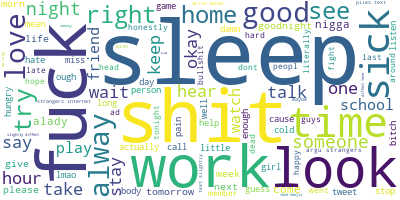} }}%
    \subfloat[\centering A word cloud for Insomnia. ]{{\includegraphics[width=4.3cm]{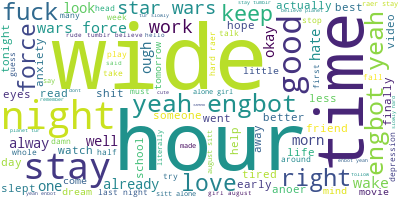} }}%
    
    \vspace{0.1cm}  
    \subfloat[\centering A word cloud for depression mood. ]{{\includegraphics[width=4.3cm]{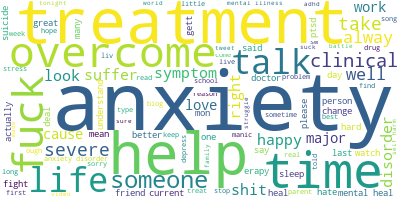} }}%
     \vspace{0.1cm}
    \caption{A word cloud depicting the most influencing symptoms.}%
    \label{fig_wordcloud}%
\end{figure}
\newpage
\subsection{Case Study}

In order to illustrate the importance of MDHAN for explaining
depression detection results, we have used an example of a depressed user to show the tweets captured by MDAHN in Figure \ref{attention_score}. The figure demonstrates that the attention map gives higher weights to explainable depression tweets; for instance, the tweet \emph{"One in four experience mental illness ... "} gained the highest attention score among all the user tweets. Moreover, MDAHN can give higher weights to explainable tweets than those interfering and unrelated tweets, which can help select more related tweets and to be a more important feature to detect the depressed user. 

To further investigate the five most influencing symptoms among depressed users, we collected all the tweets associated with these symptoms. Then we created a tag cloud \cite{viegas2008timelines} for each of these five symptoms, to determine what are the frequent words and importance that related to each symptom as shown in Figure \ref{fig_wordcloud} where larger font words are relatively more important than rest in the same cloud representation. This cloud gives us an overview of all the words that occur most frequently within each of these five symptoms.

\section{Conclusion}
In this paper, we propose a new model for detecting depressed user through social media analysis by extracting features from the user behaviour and the user’s online timeline (posts). We have used a real-world data set for depressed and non-depressed users and applied them in our model. We have proposed a hybrid model which is characterised by introducing an interplay between Multi-Modal Depression Detection Hierarchical Attention Network \textbf{MDHAN}. We assign the multi-modalities attribute which represents the user behaviour into the MLP and user timeline posts into HAN to calculate each tweet and words' importance, and capture semantic sequence features from the user timelines (posts). Our model shows that by training this hybrid network improves classification performance and identifies depressed users outperforming other strong methods and ensures adequate evidence to explain the prediction.


\printcredits

\bibliographystyle{unsrt}
\bibliography{MDHAN} 

\begin{thebibliography}{10}

\bibitem{american2013diagnostic}
American~Psychiatric Association et~al.
\newblock {\em Diagnostic and statistical manual of mental disorders
  (DSM-5{\textregistered})}.
\newblock American Psychiatric Pub, 2013.

\bibitem{metzl2015mental}
Jonathan~M Metzl and Kenneth~T MacLeish.
\newblock Mental illness, mass shootings, and the politics of american
  firearms.
\newblock {\em American journal of public health}, 105(2):240--249, 2015.

\bibitem{zafar2020survey}
Aqsa Zafar and Sanjay Chitnis.
\newblock Survey of depression detection using social networking sites via data
  mining.
\newblock In {\em 2020 10th International Conference on Cloud Computing, Data
  Science \& Engineering (Confluence)}, pages 88--93. IEEE, 2020.

\bibitem{WHO3}
World Health~Organization WHO.
\newblock Depression.
\newblock {\em http://www.who.int/mediacentre/factsheets/fs369/en/}, 2015.

\bibitem{rissola2020beyond}
Esteban~Andr{\'e}s R{\'\i}ssola, Mohammad Aliannejadi, and Fabio Crestani.
\newblock Beyond modelling: Understanding mental disorders in online social
  media.
\newblock In {\em European Conference on Information Retrieval}, pages
  296--310. Springer, 2020.

\bibitem{zou2020depression}
Maria~Li Zou, Mandy~Xiaoyang Li, and Vincent Cho.
\newblock Depression and disclosure behavior via social media: A study of
  university students in china.
\newblock {\em Heliyon}, 6(2):e03368, 2020.

\bibitem{hawn2009take}
Carleen Hawn.
\newblock Take two aspirin and tweet me in the morning: how twitter, facebook,
  and other social media are reshaping health care.
\newblock {\em Health affairs}, 28(2):361--368, 2009.

\bibitem{neuhauser2003rethinking}
Linda Neuhauser and Gary~L Kreps.
\newblock Rethinking communication in the e-health era.
\newblock {\em Journal of Health Psychology}, 8(1):7--23, 2003.

\bibitem{scanfeld2010dissemination}
Daniel Scanfeld, Vanessa Scanfeld, and Elaine~L Larson.
\newblock Dissemination of health information through social networks: Twitter
  and antibiotics.
\newblock {\em American journal of infection control}, 38(3):182--188, 2010.

\bibitem{prier2011identifying}
Kyle~W Prier, Matthew~S Smith, Christophe Giraud-Carrier, and Carl~L Hanson.
\newblock Identifying health-related topics on twitter.
\newblock In {\em International conference on social computing,
  behavioral-cultural modeling, and prediction}, pages 18--25. Springer, 2011.

\bibitem{park2012depressive}
Minsu Park, Chiyoung Cha, and Meeyoung Cha.
\newblock Depressive moods of users portrayed in twitter.
\newblock In {\em Proceedings of the ACM SIGKDD Workshop on healthcare
  informatics (HI-KDD)}, volume 2012, pages 1--8, 2012.

\bibitem{bathina2020depressed}
Krishna~C Bathina, Marijn~ten Thij, Lorenzo Lorenzo-Luaces, Lauren~A Rutter,
  and Johan Bollen.
\newblock Depressed individuals express more distorted thinking on social
  media.
\newblock {\em arXiv preprint arXiv:2002.02800}, 2020.

\bibitem{zogan2021depression}
Hamad Zogan, Imran Razzak, Shoaib Jameel, and Guandong Xu.
\newblock Depressionnet: Learning multi-modalities with user post summarization
  for depression detection on social media.
\newblock {\em Proceedings of the 44rd International ACM SIGIR Conference on
  Research and Development in Information Retrieval}, 2021.

\bibitem{kumar2019anxious}
Akshi Kumar, Aditi Sharma, and Anshika Arora.
\newblock Anxious depression prediction in real-time social data.
\newblock {\em Available at SSRN 3383359}, 2019.

\bibitem{lebret2015rehabilitation}
R{\'e}mi Lebret and Ronan Collobert.
\newblock Rehabilitation of count-based models for word vector representations.
\newblock In {\em International Conference on Intelligent Text Processing and
  Computational Linguistics}, pages 417--429. Springer, 2015.

\bibitem{de2013predicting}
Munmun De~Choudhury, Michael Gamon, Scott Counts, and Eric Horvitz.
\newblock Predicting depression via social media.
\newblock In {\em Seventh international AAAI conference on weblogs and social
  media}, 2013.

\bibitem{hu2015predicting}
Quan Hu, Ang Li, Fei Heng, Jianpeng Li, and Tingshao Zhu.
\newblock Predicting depression of social media user on different observation
  windows.
\newblock In {\em 2015 IEEE/WIC/ACM International Conference on Web
  Intelligence and Intelligent Agent Technology (WI-IAT)}, volume~1, pages
  361--364. IEEE, 2015.

\bibitem{tsugawa2015recognizing}
Sho Tsugawa, Yusuke Kikuchi, Fumio Kishino, Kosuke Nakajima, Yuichi Itoh, and
  Hiroyuki Ohsaki.
\newblock Recognizing depression from twitter activity.
\newblock In {\em Proceedings of the 33rd annual ACM conference on human
  factors in computing systems}, pages 3187--3196. ACM, 2015.

\bibitem{resnik-etal-2015-beyond}
Philip Resnik, William Armstrong, Leonardo Claudino, Thang Nguyen, Viet-An
  Nguyen, and Jordan Boyd-Graber.
\newblock Beyond {LDA}: Exploring supervised topic modeling for
  depression-related language in twitter.
\newblock In {\em Proceedings of the 2nd Workshop on Computational Linguistics
  and Clinical Psychology: From Linguistic Signal to Clinical Reality}, pages
  99--107, Denver, Colorado, June 5 2015. Association for Computational
  Linguistics.

\bibitem{shen2017depression}
Guangyao Shen, Jia Jia, Liqiang Nie, Fuli Feng, Cunjun Zhang, Tianrui Hu,
  Tat-Seng Chua, and Wenwu Zhu.
\newblock Depression detection via harvesting social media: A multimodal
  dictionary learning solution.
\newblock In {\em IJCAI}, pages 3838--3844, 2017.

\bibitem{soton423226}
Tiancheng Shen, Jia Jia, Guangyao Shen, Fuli Feng, Xiangnan He, Huanbo Luan,
  Jie Tang, Thanassis Tiropanis, Tat~Seng Chua, and Wendy Hall.
\newblock Cross-domain depression detection via harvesting social media.
\newblock In {\em Proceedings of the 27th International Joint Conference on
  Artificial Intelligence, IJCAI 2018}, volume 2018-July, pages 1611--1617.
  International Joint Conferences on Artificial Intelligence, July 2018.

\bibitem{ramirez2018early}
Diana Ram{\'i}rez-Cifuentes, Marc Mayans, and Ana Freire.
\newblock Early risk detection of anorexia on social media.
\newblock In {\em International Conference on Internet Science}, pages 3--14.
  Springer, 2018.

\bibitem{yazdavar2017semi}
Amir~Hossein Yazdavar, Hussein~S Al-Olimat, Monireh Ebrahimi, Goonmeet Bajaj,
  Tanvi Banerjee, Krishnaprasad Thirunarayan, Jyotishman Pathak, and Amit
  Sheth.
\newblock Semi-supervised approach to monitoring clinical depressive symptoms
  in social media.
\newblock In {\em Proceedings of the 2017 IEEE/ACM International Conference on
  Advances in Social Networks Analysis and Mining 2017}, pages 1191--1198. ACM,
  2017.

\bibitem{wang2017detecting}
Tao Wang, Markus Brede, Antonella Ianni, and Emmanouil Mentzakis.
\newblock Detecting and characterizing eating-disorder communities on social
  media.
\newblock In {\em Proceedings of the Tenth ACM International Conference on Web
  Search and Data Mining}, pages 91--100. ACM, 2017.

\bibitem{wongkoblap2019modeling}
Akkapon Wongkoblap, Miguel~A Vadillo, and Vasa Curcin.
\newblock Modeling depression symptoms from social network data through
  multiple instance learning.
\newblock {\em AMIA Summits on Translational Science Proceedings}, 2019:44,
  2019.

\bibitem{devlin2018bert}
Jacob Devlin, Ming-Wei Chang, Kenton Lee, and Kristina Toutanova.
\newblock \uppercase{Bert}: Pre-training of deep bidirectional transformers for
  language understanding.
\newblock {\em arXiv preprint arXiv:1810.04805}, 2018.

\bibitem{chiu2020multimodal}
Chun~Yueh Chiu, Hsien~Yuan Lane, Jia~Ling Koh, and Arbee~LP Chen.
\newblock Multimodal depression detection on instagram considering time
  interval of posts.
\newblock {\em Journal of Intelligent Information Systems}, pages 1--23, 2020.

\bibitem{lin2020sensemood}
Chenhao Lin, Pengwei Hu, Hui Su, Shaochun Li, Jing Mei, Jie Zhou, and Henry
  Leung.
\newblock Sensemood: Depression detection on social media.
\newblock In {\em Proceedings of the 2020 International Conference on
  Multimedia Retrieval}, pages 407--411, 2020.

\bibitem{zheng2020graph}
Wenbo Zheng, Lan Yan, Chao Gou, and Fei-Yue Wang.
\newblock Graph attention model embedded with multi-modal knowledge for
  depression detection.
\newblock In {\em 2020 IEEE International Conference on Multimedia and Expo
  (ICME)}, pages 1--6. IEEE, 2020.

\bibitem{xezonaki2020affective}
Danai Xezonaki, Georgios Paraskevopoulos, Alexandros Potamianos, and Shrikanth
  Narayanan.
\newblock Affective conditioning on hierarchical networks applied to depression
  detection from transcribed clinical interviews.
\newblock {\em arXiv preprint arXiv:2006.08336}, 2020.

\bibitem{wolohan2020estimating}
JT~Wolohan.
\newblock Estimating the effect of covid-19 on mental health: Linguistic
  indicators of depression during a global pandemic.
\newblock In {\em Proceedings of the 1st Workshop on NLP for COVID-19 at ACL
  2020}, 2020.

\bibitem{mikolov2017advances}
Tomas Mikolov, Edouard Grave, Piotr Bojanowski, Christian Puhrsch, and Armand
  Joulin.
\newblock Advances in pre-training distributed word representations.
\newblock {\em arXiv preprint arXiv:1712.09405}, 2017.

\bibitem{shrestha2020multi}
Anu Shrestha, Edoardo Serra, and Francesca Spezzano.
\newblock Multi-modal social and psycho-linguistic embedding via recurrent
  neural networks to identify depressed users in online forums.
\newblock {\em NetMAHIB}, 9(1):22, 2020.

\bibitem{trotzek2018utilizing}
Marcel Trotzek, Sven Koitka, and Christoph~M Friedrich.
\newblock Utilizing neural networks and linguistic metadata for early detection
  of depression indications in text sequences.
\newblock {\em IEEE Transactions on Knowledge and Data Engineering}, 2018.

\bibitem{farruque2019augmenting}
Nawshad Farruque, Osmar Zaiane, and Randy Goebel.
\newblock Augmenting semantic representation of depressive language: From
  forums to microblogs.
\newblock In {\em Joint European Conference on Machine Learning and Knowledge
  Discovery in Databases}, pages 359--375. Springer, 2019.

\bibitem{faruqui2014retrofitting}
Manaal Faruqui, Jesse Dodge, Sujay~K Jauhar, Chris Dyer, Eduard Hovy, and
  Noah~A Smith.
\newblock Retrofitting word vectors to semantic lexicons.
\newblock {\em arXiv preprint arXiv:1411.4166}, 2014.

\bibitem{gong2017topic}
Yuan Gong and Christian Poellabauer.
\newblock Topic modeling based multi-modal depression detection.
\newblock In {\em Proceedings of the 7th Annual Workshop on Audio/Visual
  Emotion Challenge}, pages 69--76, 2017.

\bibitem{peng2019multi}
Zhichao Peng, Qinghua Hu, and Jianwu Dang.
\newblock Multi-kernel svm based depression recognition using social media
  data.
\newblock {\em International Journal of Machine Learning and Cybernetics},
  10(1):43--57, 2019.

\bibitem{article}
Krystian Horecki and Jacek Mazurkiewicz.
\newblock Natural language processing methods used for automatic prediction
  mechanism of related phenomenon.
\newblock {\em Lecture Notes in Artificial Intelligence (Subseries of Lecture
  Notes in Computer Science)}, 9120:13--24, 06 2015.

\bibitem{8389299}
M.~{Deshpande} and V.~{Rao}.
\newblock Depression detection using emotion artificial intelligence.
\newblock In {\em 2017 International Conference on Intelligent Sustainable
  Systems (ICISS)}, pages 858--862, Dec 2017.

\bibitem{novak2015sentiment}
Petra~Kralj Novak, Jasmina Smailovi{\'c}, Borut Sluban, and Igor Mozeti{\v{c}}.
\newblock Sentiment of emojis.
\newblock {\em PloS one}, 10(12):e0144296, 2015.

\bibitem{blei2003latent}
David~M Blei, Andrew~Y Ng, and Michael~I Jordan.
\newblock Latent dirichlet allocation.
\newblock {\em Journal of machine Learning research}, 3(Jan):993--1022, 2003.

\bibitem{teh2005sharing}
Yee~W Teh, Michael~I Jordan, Matthew~J Beal, and David~M Blei.
\newblock Sharing clusters among related groups: Hierarchical dirichlet
  processes.
\newblock In {\em Advances in neural information processing systems}, pages
  1385--1392, 2005.

\bibitem{yang2016hierarchical}
Zichao Yang, Diyi Yang, Chris Dyer, Xiaodong He, Alex Smola, and Eduard Hovy.
\newblock Hierarchical attention networks for document classification.
\newblock In {\em Proceedings of the 2016 conference of the North American
  chapter of the association for computational linguistics: human language
  technologies}, pages 1480--1489, 2016.

\bibitem{pennington2014glove}
Jeffrey Pennington, Richard Socher, and Christopher~D Manning.
\newblock Glove: Global vectors for word representation.
\newblock In {\em Proceedings of the 2014 conference on empirical methods in
  natural language processing (EMNLP)}, pages 1532--1543, 2014.

\bibitem{karmen2015screening}
Christian Karmen, Robert~C Hsiung, and Thomas Wetter.
\newblock Screening internet forum participants for depression symptoms by
  assembling and enhancing multiple nlp methods.
\newblock {\em Computer methods and programs in biomedicine}, 120(1):27--36,
  2015.

\bibitem{ng2002discriminative}
Andrew~Y Ng and Michael~I Jordan.
\newblock On discriminative vs. generative classifiers: A comparison of
  logistic regression and naive bayes.
\newblock In {\em Advances in neural information processing systems}, pages
  841--848, 2002.

\bibitem{cho-etal-2014-learning}
Kyunghyun Cho, Bart van Merri{\"e}nboer, Caglar Gulcehre, Dzmitry Bahdanau,
  Fethi Bougares, Holger Schwenk, and Yoshua Bengio.
\newblock Learning phrase representations using {RNN} encoder{--}decoder for
  statistical machine translation.
\newblock In {\em Proceedings of the 2014 Conference on Empirical Methods in
  Natural Language Processing ({EMNLP})}, pages 1724--1734, Doha, Qatar,
  October 2014. Association for Computational Linguistics.

\bibitem{kim-2014-convolutional}
Yoon Kim.
\newblock Convolutional neural networks for sentence classification.
\newblock In {\em Proceedings of the 2014 Conference on Empirical Methods in
  Natural Language Processing ({EMNLP})}, pages 1746--1751, Doha, Qatar,
  October 2014. Association for Computational Linguistics.

\bibitem{Kingma2015AdamAM}
Diederik~P. Kingma and Jimmy Ba.
\newblock Adam: A method for stochastic optimization.
\newblock {\em CoRR}, abs/1412.6980, 2015.

\bibitem{viegas2008timelines}
Fernanda~B Vi{\'e}gas and Martin Wattenberg.
\newblock Timelines tag clouds and the case for vernacular visualization.
\newblock {\em interactions}, 15(4):49--52, 2008.

\end{thebibliography}

\section{Appendix A}
\label{appendix:symptoms}
List of depression symptoms as per DSM-IV:
\begin{enumerate}
  \item Depressed mood.
  \item iminished interest.
  \item Weight or appetite change
  \item Insomnia, hypersomnia.
  \item Psychomotor retardation, psychomotor impairment.
  \item Fatigue or loss of energy
  \item Feelings worthlessness, guilt.
  \item Diminished ability to think, indecisiveness.
  \item Suicidal tendency.
\end{enumerate}


\section{Appendix B}
\label{appendix:antidepressant}

List of antidepressant medicine names

\begin{table*}[]
\begin{tabular}{llllll}
Citalopram     & Celexa           & Cipramil          & Escitalopram  & Lexapro       & Cipralex        \\
Fluoxetine     & Prozac           & Sarafem           & Fluvoxamine   & Luvox         & Faverin         \\
Paroxetine     & Paxil            & Seroxat           & Sertraline    & Zoloft        & Lustral         \\
Desvenlafaxine & Pristiq          & Duloxetine        & Cymbalta      & Levomilnac.   & Fetzima         \\
Milnacipran    & Ixel             & Savella           & Venlafaxine   & Effexor       & Vilazodone      \\
Viibryd        & Vortioxetine     & Trintellix        & Nefazodone    & Dutonin       & Nefadar         \\
Serzone        & Trazodone        & Desyrel           & Atomoxetine   & Strattera     & Reboxetine      \\
Edronax        & Teniloxazine     & Lucelan           & Metatone      & Viloxazine    & Vivalan         \\
Bupropion      & Wellbutrin       & Amitriptyline     & Elavil        & Endep         & Trifluoperazine \\
Amioxid        & Ambivalon        & Equilibrin        & Clomipramine  & Anafranil     & Desipramine     \\
Norpramin      & Pertofrane       & Dibenzepin        & Noveril       & Victoril      & Dimetacrine     \\
Istonil        & Dosulepin        & Prothiaden        & Doxepin       & Adapin        & Sinequan        \\
Imipramine     & Tofranil         & Lofepramine       & Lomont        & Gamanil       & Melitracen      \\
Dixeran        & Melixeran        & Trausabun         & Nitroxazepine & Sintamil      & Nortriptyline   \\
Pamelor        & Aventyl          & Noxiptiline       & Agedal        & Elronon       & Nogedal         \\
Opipramol      & Insidon          & Pipofezine        & Azafen        & Azaphen       & Protriptyline   \\
Vivactil       & Trimipramine     & Surmontil         & Amoxapine     & Asendin       & Maprotiline     \\
Ludiomil       & Mianserin        & Tolvon            & Mirtazapine   & Remeron       & Setiptiline     \\
Tecipul        & Mianserin        & mirtazapine       & setiptiline   & Isocarboxazid & Marplan         \\
Phenelzine     & Nardil           & Tranylcyp.        & Parnate       & Selegiline    & Eldepryl        \\
Zelapar        & Emsam            & Caroxazone        & Surodil       & Timostenil    & Metralindole    \\
Inkazan        & Moclobemide      & Aurorix           & Manerix       & Pirlindole    & Pirazidol       \\
Toloxatone     & Humoryl          & Eprobemide        & Befol         & Minaprine     & Brantur         \\
Cantor         & Bifemelane       & Alnert            & Celeport      & Agomelatine   & Valdoxan        \\
Esketamine     & Spravato         & Ketamine          & Ketalar       & Tandospirone  & Sediel          \\
Tianeptine     & Stablon          & Coaxil            & Indeloxazine  & Elen          & Noin            \\
Medifoxamine   & Clédial          & Gerdaxyl          & Oxaflozane    & Conflictan    & Pivagabine      \\
Tonerg         & Ademetionine     & Aurorix           & SAMe          & Heptral       & Transmetil      \\
Samyl          & Hypericum   per. & St.   John’s Wort & SJW           & Jarsin        & Kira            \\
Movina         & Oxitriptan       & Kira              & 5-HTP         & Cincofarm     & Levothym        \\
Triptum        & Rubidium   chl.  & Rubinorm          & Tryptophan    & Tryptan       & Optimax         \\
Aminomine      & Magnesium        & Noveril           & Solian        & Aripiprazole  & Abilify         \\
Brexpiprazole  & Rexulti          & Lurasidone        & Latuda        & Olanzapine    & Zyprexa         \\
Quetiapine     & Seroquel         & Risperidone       & Risperdal     & Buspirone     & Buspar          \\
Lithium        & Eskalith         & Lithobid          & Modafinil     & Thyroxine     & Triiodoth.      \\
Minocycline    & Amitriptyline    & chlordiaz.        & Limbitrol     & Parmodalin    & Aurorix         \\
Perphenazine   & Etafron          & Flupentixol       & melitracen    & Deanxit       & Surodil        
\end{tabular}
\end{table*}

\end{document}